%% file: 2026-seaa-isia-af.tex
\documentclass[runningheads]{llncs}
\usepackage[T1]{fontenc}
\usepackage[hidelinks]{hyperref}
\usepackage{graphicx}
\usepackage{glossaries}
\usepackage{xcolor}
\usepackage{tikz}
\usepackage{booktabs}
\usepackage{multicol}
\usepackage{multirow}
\usepackage{listings}
\input{acronyms}

\newcommand{\framework}{\textit{ISIA-AF}}

\newcommand\copyrighttext{%
  \footnotesize This preprint has not undergone peer review or any post-submission improvements or corrections. The Version of Record of this contribution is published in Software Engineering and Advanced Applications, SEAA 2026. Lecture Notes in Computer Science, vol 16863, and is available online at \url{https://doi.org/10.1007/978-3-032-36590-3_2}.
}
\newcommand\copyrightnotice{%
  \begin{tikzpicture}[remember picture, overlay]
    \node[anchor=south, yshift=10pt] at (current page.south)
      {\fbox{\parbox{\dimexpr\textwidth-\fboxsep-\fboxrule\relax}{\copyrighttext}}};
  \end{tikzpicture}%
}

\begin{document}
\title{ISIA-AF: Orchestrating Reproducible Attacks and Multi-Source Data Collection for OT Systems}
\titlerunning{ISIA-AF: An Attack Framework for OT Systems}
\author{Stefan Manfred Haratzmüller\inst{1} \and Thomas Rosenstatter\inst{1}%
\and \\ Olaf Saßnick\inst{1,2}%
\and Dalibor Sain\inst{1}
\and Stefan Huber\inst{1}%
}
\authorrunning{S. M. Haratzmüller et al.}
\institute{Josef Ressel Centre for Intelligent and Secure Industrial Automation, Salzburg University of Applied Sciences, 5412 Puch/Salzburg, Austria
\email{\texttt{[firstname.lastname]}@fh-salzburg.ac.at}\\\and
Paris Lodron University Salzburg, 5020 Salzburg, Austria}
\maketitle              %
\copyrightnotice
\begin{abstract}
\input{abstract}
\end{abstract}
\section{Introduction}
\label{sec:introduction}
\gls{ot} requires increased security as threats to these systems have increased tremendously in recent years. 
This is partly due to the growing interconnectivity and complexity of these systems, which leads to larger attack surfaces. 
For instance, in 2017 Triton surfaced, the first attempt against safety systems targeting Triconex SIS controllers~\cite{Makrakis2021IndustrialCriticalInfaSecurity}.

To improve security, datasets are essential for training and evaluating security mechanisms. 
For this, anomaly and \glspl{ids} demand realistic datasets to increase their detection rate which in turn need to consist of a wide range of attack scenarios and diversity of technologies.
Despite this, security datasets for \gls{ot} systems, especially those including newer, state-of-the-art protocols like \gls{opcua}, are scarce. 
Only two datasets~\cite{pinto2020m2m,Radhakrishnan2025Unsupervised} for security research use \gls{opcua}, and both are based on networks of Raspberry Pi devices running a Python \gls{opcua} implementation.
Of these two, only Pinto~\cite{pinto2020m2m} made the dataset publicly available.
X-IIoTID~\cite{AlHawawreh2022XIIoTID}, one of the most extensive \gls{ot} datasets, includes network traffic, host logs, system resources, and \gls{ids} alerts, with support for protocols such as Modbus, MQTT, and CoAP.
IoTForge Pro~\cite{kumar2025} introduces a security testbed for IIoT environments that generates also the ForgeIIOT dataset. %
While IoTForge Pro demonstrates a structured approach to dataset generation, the attack scripts and automation details are not publicly available, leaving the methodology difficult to reproduce or extend.

Conti et al.~\cite{Conti2021SurveyICSTestbeds} identify the most important dataset properties as \emph{reproducibility}, \emph{realism}, and \emph{extensibility}. %
Goldschmidt and Chudák~\cite{goldschmidt2025} %
criticise that security datasets are typically published without the complete documentation for full reproducibility.
For instance, X-IIoTID~\cite{AlHawawreh2022XIIoTID} is publicly available, 
but details of the automation used for traffic generation and attack scenarios are not released.
As such, extending the dataset with more recent attacks or building new datasets is more challenging, and the successful path and tooling required to create high-quality datasets remain unclear.

While general attack simulation and automation frameworks~\cite{MITRECaldera,AtomicRedTeam} are available, 
they are typically not designed for reproducible dataset generation. %
MITRE Caldera~\cite{MITRECaldera}, for example, focuses on mimicking real attackers by adding stochasticity and employing a model based on abilities and facts,
which makes it difficult to execute reproducible attack chains.
General-purpose automation frameworks like Ansible~\cite{Ansible} exist, but they are primarily designed for sequential task execution on remote hosts. 
Although such frameworks can orchestrate distributed workflows, the SSH-based execution of Ansible adds overhead and may introduce timing artefacts.
Ultimately, LCM~\cite{huang2010lcm}, a lightweight communication framework, was selected for its flexibility while also minimising overhead.

\paragraph{Contributions.}~In this work, we present {\it ISIA-AF}, an attack framework that enables the execution of a wide range of attacks on industrial systems.
The framework is used not only to coordinate attack scenarios, but also to automate the generation of a security dataset containing data from both network traffic and operational data from the industrial process, thereby \textit{enriching the dataset}.
Overall, \framework{} (i) enables reproducible attack runs, (ii) automates dataset generation, (iii) enriches the dataset with multiple data sources, (iv) provides a modular and extensible architecture for the development of new attacks, and (v) is publicly available\footnote{\url{https://github.com/JRC-ISIA/isia-attack-framework}}.
We demonstrate its functionality by deploying it on an industrial testbed that is representative of a real-world industrial environment.%

\section{Attack Framework Design}
The design of \framework{} follows Wieringa's \gls{dsr} methodology~\cite{Wieringa2014DesignScience}, which structures artefact construction and evaluation around a \textit{design cycle} of three iterative activities: (i)~\textit{problem investigation}, identifying requirements from the security and dataset needs of \gls{ot} environments; (ii)~\textit{treatment design}, specifying the architecture and features of \framework{}; and (iii)~\textit{treatment validation}, demonstrating the framework's utility through a case study on the \gls{isia} testbed.

The central design problem is the lack of systematic and automated means of executing, coordinating, and documenting attack scenarios on \gls{ot} systems to generate reproducible, enriched security datasets, a gap highlighted in literature~\cite{Conti2021SurveyICSTestbeds,goldschmidt2025}.
\framework{} addresses this through a modular and server/client architecture that captures multi-type data.

\paragraph{Objectives.}~Table~\ref{tab:requirements} provides a list of the functional and non-functional requirements derived from literature~\cite{Conti2021SurveyICSTestbeds,goldschmidt2025} and discussions with the stakeholders at the \gls{isia} research centre.

\begin{table}[t]
	\centering
	\caption{Functional and non-functional requirements for the \framework{} derived from literature and discussions with stakeholders.}\label{tab:requirements}
 	\renewcommand{\arraystretch}{1.1}\footnotesize
	\begin{tabular}{p{.08\linewidth}p{.89\linewidth}}
	\toprule
	{R.\#} &  {Requirement Description}\\
	\midrule
	\multicolumn{2}{l}{\textit{Functional Requirements}}\\\cline{1-1}
        FR.1 & Load attack definition using a structured data format (e.g., JSON or YAML). \\
        FR.2 & Execution without an attack must be possible (``no-attack'' scenario). \\
        FR.3 & Execution of the scenario can be triggered manually or by a schedule. \\
        FR.4 & Record traffic on various OSI layers. \\
        FR.5 & Log metadata such as timestamps, attack names, and success/failure flags. \\
        FR.6 & Extract relevant features from recorded traffic for dataset creation. \\
        FR.7 & Label captured data as \textit{under attack} or to a normal system state. \\
        FR.8 & Provide reliable persistent data storage for later use and archival purposes.\\
	\multicolumn{2}{l}{\textit{Non-Functional Requirements}}\\\cline{1-1}
        NF.1 & The framework must be modular to allow adaptability and easy extension with new features or components. \\
        NF.2 & Coordination and control must remain centralized.\\
        NF.3 & Attack clients must allow new attack types that can be implemented and integrated independently. \\
		NF.4 & Communication and coordination with minimal communication overhead.\\
		NF.5 & Provide temporarily (buffer) storage during processing. \\

	\bottomrule
	\end{tabular}

\end{table}

To achieve these objectives, such as modular design, centralised control, and minimal overhead, the framework consists of the following component types: \textit{attack orchestrator}, \textit{attack clients}, \textit{capture modules}, \textit{logger}, \textit{feature extractor}, and \textit{storage transfer} module. 
The orchestrator is responsible for the centralized control and monitoring of the framework.
The attack clients perform the timed execution of attacks, and the capture clients record network traffic and system response. 
The logger and feature extractor are responsible for logging attack activities and system responses, and data enrichment respectively.
The storage transfer is used to archive the logs centrally.

\paragraph{Placement within the Testbed.}\label{sec:placement}

The attack framework is to be placed in the ISIA testbed, which is a HIL realisation that simulates the physical process of three~\glspl{imm} and robot arms that lift the produced items on a conveyor belt. 
The control of the \glspl{imm} is performed on \glspl{plc} by B\&R Automation and can be manipulated and monitored with a \gls{scada} server by COPA-DATA.

\begin{figure}[!htb]
	\centering
	\includegraphics[width=.95\linewidth]{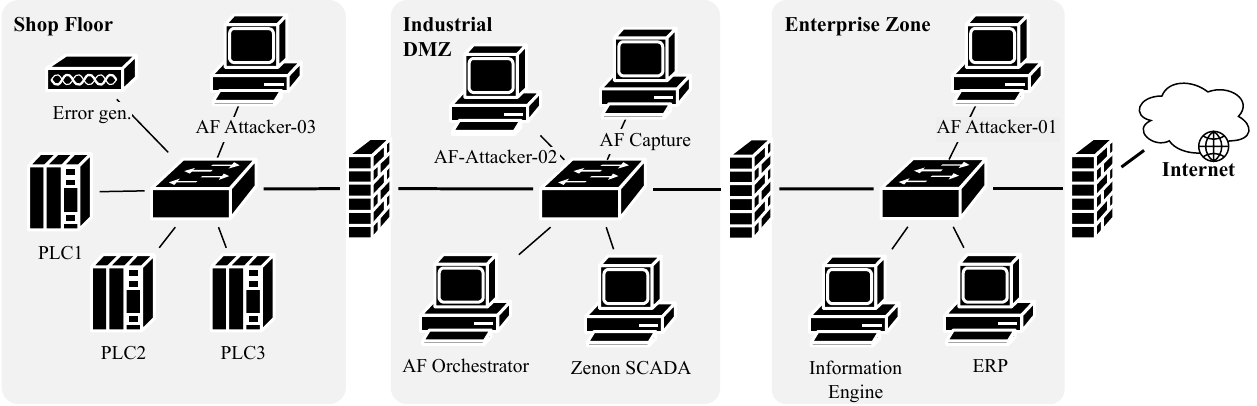 }
	\caption{Network view of the ISIA testbed with the attack framework. Attacker clients are located in each level of the industrial network, and the orchestrator and capture components are located in the industrial \gls{dmz}.}\label{fig:testbed}
\end{figure}

Figure~\ref{fig:testbed} details the network view of the ISIA testbed split into three segments following IEC~62443~\cite{iec62443-3.2} recommendations.
The leftmost segment is the \textit{shop floor}, which contains the three~\glspl{imm} and the \glspl{plc} (PLC~1-4) for controlling the production process. 
The simulation of the robot arms is performed by automation PCs (KUKA~1-4), each for one arm respectively. 
The mid-section is the \textit{Industrial \gls{dmz}} comprising the \gls{scada} system and a security server that provides additional security services.
The \textit{enterprise zone} can be found in the rightmost segment and consists of an ERP system and an information engine tasked with monitoring the operational state of the OT system. Typically, this zone is also connected to the internet.

As the framework must be modular so that various attacks can be performed (\textit{NF.3}), it is necessary to place attack clients in each network segment of the testbed. 
This distributed placement ensures that intra-segment attacks (e.g., lateral movement, ARP spoofing, or \gls{dos} within a single zone) as well as cross-segment attacks (e.g., man-in-the-middle traversing zone boundaries) can be  reproduced without constraints.

The placement of the attack orchestrator and the capture module can be adapted, but we suggest placing these components in the industrial \gls{dmz}.
The industrial \gls{dmz} is chosen due to its location, representing the proxy for all network traffic between the \gls{it} and \gls{ot} zones.
In addition, the capture module is connected to mirrored switchports, which allows it to capture all network traffic.

\subsection{Architecture}\label{sec:architecture}

All components communicate via LCM~\cite{huang2010lcm}, a lightweight publish/subscribe middleware that satisfies the minimal-overhead communication requirement (\textit{NF.4}) through UDP multicasts. 
The components subscribe to their respective channels and may further filter messages by their unique identifier (e.g., \texttt{attacker-01}).

\paragraph{Attack Orchestrator:}
The orchestrator has sole control of the components and issues timed commands to all registered clients, and aggregates metadata (\textit{FR.5}) in logs received through the logger module.
It provides a command line interface, but also allows loading attack scenario definitions encoded in \textit{JSON} (\textit{FR.1}) to either trigger them manually or schedule them for automated execution (\textit{NF.2, FR.3}).
Internally, it parses the scenario definition into an ordered list specifying the target attack client, its parameters, and an ISO~8601 timestamp or offset, and broadcasts them to the relevant clients over an LCM channel.
Upon completion of each directive, clients publish a status message that the orchestrator uses to update its execution log, recording success or failure flags alongside timestamps and further responses (\textit{FR.5}).
A ``no-attack'' scenario, in which the framework runs but no attack is issued, is also supported as an execution mode (\textit{FR.2}), thereby enabling the generation and recording of a benign baseline.

\paragraph{Attack Client:}
On receipt of a directive, the attack client dynamically loads and invokes the specified \emph{attack script} (\textit{NF.3}) enabling new attack types to be integrated without modifying the client itself (\textit{NF.3}).
The number of attack clients is variable and allows placement in any network segment. For instance, Figure~\ref{fig:testbed} illustrates three clients (\texttt{AF Attacker-01} through \texttt{AF Attacker-03}) deployed in the enterprise zone, industrial \gls{dmz}, and shop floor, respectively.

\paragraph{Capture Module:}
The capture module is a dedicated node connected to mirrored switchports in the industrial \gls{dmz}, granting it passive visibility into all inter-zone traffic without injecting additional frames into the network (\textit{NF.4}).
It records traffic based on variable filters at the network interface level, and stores the raw \texttt{.pcap} files on a network share (\textit{NF.5}) for later processing by the attack orchestrator.

\paragraph{Logger Module:}
The labelling of all data records, i.e., operational data (\gls{opcua} data) and network traffic (\textit{FR.4}), is driven by the execution log produced by the orchestrator (\textit{FR.7}). %
Execution logs contain the start and end times of attacks and normal operations, as well as outputs and errors occurring during the execution of the attack scripts. 

\paragraph{Feature Extractor:}
Temporal alignment is achieved through time synchronisation across all testbed nodes.
This allows the feature extractor to query the operational data database using the start and end times of the network traffic recordings.
The feature extractor then computes and stores the traffic features as flows similarly to~\cite{AlHawawreh2022XIIoTID}, and saves the operational data according to the \gls{opcua} data model. 
The source of the operational data is called \textit{information engine} and is further described in~\cite{HHH23}.

For labelling data, the feature extractor uses the execution log to annotate both network traffic and operational data (\textit{FR.6}).
The current version of \framework{} applies the label \textit{``under attack''} to all data during attack periods.
Future work will extend labelling via logger functions to include more granular categories such as {\it attack traffic}, {\it system reaction/recovery}, and {\it normal traffic}, as recommended by Conti et al.~\cite{Conti2021SurveyICSTestbeds}.

\paragraph{Storage Transfer:}
Once the data is labelled, it is transferred to a persistent storage unit (\textit{FR.8}) in the form of separate JSON files: labelled network flows, labelled operational data, the execution log, and the network capture as \texttt{.pcap} file.
Keeping the logs separated was a design decision to allow for easier analysis and processing of data in case only one type is of interest for evaluating an \gls{ids}. Lastly, the locally stored files are removed.

\subsection{Attack Scenarios}
Attack scenarios are defined through tasks which include the attacks themselves, but also allow to prepare the environment through tasks such as copying new scripts to attacker devices and starting or stopping network capture. 
Listing~\ref{lst:attackdefinition} shows the properties of each action, including a unique identifier, the action type, the absolute or relative execution time, and a blocking flag that either allows to directly continue with the next action  (\texttt{blocking=false}) or waits for the current action to finish before continuing with the next one (\texttt{blocking=true}). Finally, additional action-specific parameters can be passed.

\lstdefinestyle{linefont}{
    basicstyle=\ttfamily\scriptsize,
    numbers=left,
    numberstyle=\tiny\color{gray},
    numbersep=5pt,
    breaklines=true,
    showstringspaces=false,
}

\begin{lstlisting}[style=linefont, caption={Structure of attack definitions allowing to use absolute and relative time, enabling blocking of actions and defining the targets.}, label={lst:attackdefinition}]
tasks ::= [ Action* ]
Action ::= {
  id: String,
  type: { execute_local, start_capture, stop_capture, set_alias, 	start_attack_script, start_attack, stop_attack, extract_features, store_dataset },
    targets: [ String* ],
    time: (RelativeTime | AbsoluteTime) | null,
  blocking: Boolean,
  params: { (String : Value)* }
}
\end{lstlisting}

\section{Case Study}\label{sec:casestudy}

In the following case study, we define actions for three phases: (i) setup, in which the necessary artifacts are distributed; (ii) recording, during which the attacks are performed; and (iii) post-attack, during which data recordings are copied to the feature extractor for subsequent dataset generation.

The attack scenario definition can be found in the examples~\footnote{\url{https://github.com/JRC-ISIA/isia-attack-framework/}}. 
This example is chosen to emphasise the flexibility of \framework{}, not to demonstrate a new attack vector for \gls{ot} systems. 
We consider an attacker who has already managed to get access to the poorly secured monitoring system.
From there, the attacker searches the network for \gls{opcua} servers, reads their information model to identify the \glspl{plc} used in manufacturing, and then performs a \gls{dos} attack.

\textit{Phase I (setup):} The system is prepared, i.e., files are copied to the attack agents using \textit{execute\_local} and the traffic capture (\textit{start\_capture}) is started with a network filter ignoring the UDP multicast communication and remote access to the devices used to oversee their status. 
Furthermore, the logging of the \glspl{plc} is enabled by running a script with the \textit{execute\_local} command.

\textit{Phase II (recording):} For defining this attack scenario, the attacker device located in the \gls{dmz} first executes a \texttt{nmap} scan for port 4840 to identify \gls{opcua} servers.
Using a Python \gls{opcua} client, the attacker then explores the \gls{opcua} servers to identify the \glspl{plc} used in the plastic moulding process.
Finally, the attacker performs a chunk flood attack on all identified machines utilising the \textit{opcua-exploit-framework}~\cite{claroty_opcua_exploit_framework}.

\textit{Phase III (post-attack):} The capture is stopped (\textit{stop\_capture}) and the feature extraction started. 
The logs and pcap are consequently moved to a central device (coordinator) from where a database query is sent to retrieve the operational data during the time of the recording. 
More logs, like the proprietary logs recorded by the \glspl{plc}, are also copied through \textit{execute\_local} to enrich the dataset.
Once all data is collected and labelled, the action \textit{store\_dataset} transfers the data to a network storage for archiving.

The given scenario highlights the versatility of the framework. 
Not only are attacks coordinated, but traffic capture is also orchestrated, network flows are automatically annotated with \textit{under attack} and \textit{benign}, and specific scripts can be executed to enhance the dataset with additional information.

\section{Conclusion}\label{sec:conclusion}

This paper presents \framework{}, a modular attack framework for orchestrating reproducible attack execution and automated dataset generation on \gls{ot} systems.
Following Wieringa's \gls{dsr} methodology, we derived functional and non-functional requirements from prior work and stakeholder discussions, and realised a framework comprising a centralised orchestrator, distributed attack clients, a passive capture module, a logger, and a feature extractor.
The framework addresses a key gap identified in the literature~\cite{Conti2021SurveyICSTestbeds,goldschmidt2025}: the lack of systematic tooling for generating reproducible, multi-source \gls{ot} security datasets with sufficient documentation for reproducibility.

A case study on the \gls{isia} testbed demonstrated that \framework{} supports flexible, multi-segment attack deployment with minimal communication overhead, and produces enriched datasets combining network traffic and \gls{opcua} operational data.
To the best of our knowledge, \framework{} is the first attack framework to target a real-world \gls{ot} environment using \gls{opcua} as the main communication protocol.

Future work will focus on extending the labelling scheme to include more granular categories such as attack traffic, system reaction, and normal traffic, as well as expanding the attack library with additional \gls{ot}-specific attack scenarios.

\begin{credits}
\subsubsection{\ackname} The financial support by the Austrian Federal Ministry of Economy, Energy and Tourism, the National Foundation for Research, Technology and Development and the Christian Doppler Research Association is gratefully acknowledged.

\end{credits}
\bibliographystyle{splncs04}
\bibliography{references}
\end{document}

%% file: acronyms.tex
\newacronym{dos}{DoS}{Denial of Service}
\newacronym{gds}{GDS}{Global Discovery Service}
\newacronym{https}{HTTPS}{Hypertext Transfer Protocol Secure}
\newacronym{imm}{IMM}{Injection Moulding Machine}
\newacronym{it}{IT}{Information Technology}
\newacronym{ids}{IDS}{Intrusion Detection System}
\newacronym{isia}{ISIA}{Intelligent and Secure Industrial Automation}
\newacronym{mqtt}{MQTT}{Message Queuing Telemetry Transport}
\newacronym{opcua}{OPC~UA}{Open Platform Communications Unified Architecture}
\newacronym{opc}{OPC}{Open Platform Communications}
\newacronym{ot}{OT}{Operational Technology}
\newacronym{pki}{PKI}{Public Key Infrastructure}
\newacronym{plc}{PLC}{Programmable Logic Controller}
\newacronym{scada}{SCADA}{Supervisory Control and Data Acquisition}
\newacronym{uri}{URI}{Uniform Resource Identifier}
\newacronym{dsr}{DSR}{Design Science Research}
\newacronym{dmz}{DMZ}{Demilitarized Zone}
\newacronym{mitm}{MITM}{Man-in-the-Middle}

%% file: abstract.tex
Operational Technology (OT) environments require realistic, reproducible security datasets, yet existing approaches often lack automation, multi-source data capture, and sufficient documentation for reuse. This paper presents ISIA-AF, a modular attack framework for orchestrating reproducible attack execution and automated dataset generation on industrial systems. The framework coordinates distributed attack clients, records network traffic and operational data, ultimately leading to a multi-source dataset. We derive functional and non-functional requirements from prior work and stakeholder discussions, and realise the framework following a design science research approach. A case study on the ISIA testbed, comprising a real industrial system and a simulated process, demonstrates how the framework supports centralised control, low communication overhead, and flexible deployment across network segments. The result is a practical basis for generating extensible, multi-source OT security datasets for intrusion detection research.